\documentclass[aps,superscriptaddress,amsmath,amssymb,twocolumn,showpacs,floatfix,reprint,longbibliography]{revtex4-2}

\usepackage{url}
\usepackage{bm}
\usepackage{graphicx}
\usepackage{amsmath}
\usepackage{amstext}
\usepackage{amssymb}
\usepackage{amsfonts}
\usepackage{amsbsy}
\usepackage{verbatim}
\usepackage{color}
\usepackage[colorlinks=true, urlcolor=blue, linkcolor=blue, citecolor=blue, pdftex]{hyperref}
\usepackage{multirow}
\usepackage{floatrow}
\usepackage{float}
\usepackage{gensymb}
\usepackage{textcomp}
\usepackage{enumitem}
\usepackage[version=3]{mhchem}
\usepackage{afterpage}
\usepackage{pbox}
\usepackage{makecell}
\usepackage{array,booktabs}
\usepackage{dsfont}
\usepackage{siunitx}

\newcommand{\dagga}{{\phantom{\dagger}}}

\begin{document}

\title{Gutzwiller-projected states for the $J_1$\textendash$J_2$ Heisenberg model on the kagome lattice: achievements and pitfalls}

\author{Yasir Iqbal}
\email[]{yiqbal@physics.iitm.ac.in}
\thanks{These authors contributed equally.}
\affiliation{Department of Physics and Quantum Centers in Diamond and Emerging Materials (QuCenDiEM) group, Indian Institute of Technology Madras, Chennai 600036, India}
\author{Francesco Ferrari}
\thanks{These authors contributed equally.}
\affiliation{Institute for Theoretical Physics, Goethe University Frankfurt, Max-von-Laue-Straße 1, D-60438 Frankfurt am Main, Germany}
\author{Aishwarya Chauhan}
\thanks{These authors contributed equally.}
\affiliation{Department of Physics and Quantum Centers in Diamond and Emerging Materials (QuCenDiEM) group, Indian Institute of Technology Madras, Chennai 600036, India}
\author{Alberto Parola}
\affiliation{Dipartimento di Scienza e Alta Tecnologia, Universit\`a dell'Insubria, Via Valleggio 11, I-22100 Como, Italy}
\author{Didier Poilblanc}
\affiliation{Laboratoire de Physique Th\'eorique UMR-5152, CNRS and Universit\'e de Toulouse, F-31062, 
Toulouse, France}
\author{Federico Becca}
\affiliation{Dipartimento di Fisica, Universit\`a di Trieste, Strada Costiera 11, I-34151 Trieste, Italy}

\date{\today}

\begin{abstract}
We assess the ground-state phase diagram of the $J_1$\textendash$J_2$ Heisenberg model on the kagome lattice by employing Gutzwiller-projected 
fermionic wave functions. Within this framework, different states can be represented, defined by distinct unprojected fermionic Hamiltonians 
that include hopping and pairing terms, as well as a coupling to local Zeeman fields to generate magnetic order. For $J_2=0$, the so-called 
U(1) Dirac state, in which only hopping is present (such as to generate a $\pi$-flux in the hexagons), has been shown to accurately describe 
the exact ground state [Y. Iqbal, F. Becca, S. Sorella, and D. Poilblanc, \href{https://doi.org/10.1103/PhysRevB.87.060405}{Phys. Rev. B {\bf 87}, 
060405 (2013)}; Y.-C. He, M. P. Zaletel, M. Oshikawa, and F. Pollmann, \href{https://doi.org/10.1103/PhysRevX.7.031020}{Phys. Rev. X {\bf 7}, 
031020 (2017)}]. Here, we show that its accuracy improves in presence of a small {\it antiferromagnetic} super-exchange $J_2$, leading to a 
finite region where the gapless spin liquid is stable; then, for $J_2/J_1=0.11(1)$, a first-order transition to a magnetic phase with pitch 
vector ${\bf q}=(0,0)$ is detected, by allowing magnetic order within the fermionic Hamiltonian. Instead, for small {\it ferromagnetic} values 
of $|J_2|/J_1$, the situation is more contradictory. While the U(1) Dirac state remains stable against several perturbations in the fermionic 
part (i.e., dimerization patterns or chiral terms), its accuracy clearly deteriorates on small systems, most notably on $36$ sites where exact 
diagonalization is possible. Then, upon increasing the ratio $|J_2|/J_1$, a magnetically ordered state with $\sqrt{3} \times \sqrt{3}$ periodicity 
eventually overcomes the U(1) Dirac spin liquid. Within the ferromagnetic $J_{2}$ regime, evidence is shown in favor of a first-order transition 
at $J_2/J_1=-0.065(5)$. 
\end{abstract}

\maketitle

\section{Introduction} 

The Heisenberg Hamiltonian
\begin{equation}
\hat{{\cal H}} = \sum_{i,j} J_{i,j} \mathbf{\hat{S}}_{i} \cdot \mathbf{\hat{S}}_{j},
\end{equation}
for spin-$S$ operators, $\mathbf{\hat{S}}_{i}=(\hat{S}_i^x,\hat{S}_i^y,\hat{S}_i^z)$, arranged on a crystal lattice, represents one of the pillars 
of condensed-matter physics, capturing fundamental phenomena in quantum magnetism, such as symmetry breaking with Goldstone excitations, quantum 
phase transitions, topological order, and fractionalization emerging from exotic ground states~\cite{sachdevbook,savary2016}. Particularly 
interesting are the cases with small spins (e.g., $S=1/2$) on highly-frustrated low-dimensional lattices (e.g., featuring a triangular {\it motif}), 
for which there is increasing theoretical and experimental evidence that unconventional phases, which cannot be described by standard mean-field 
approaches, may settle down at sufficiently low temperatures~\cite{savary2016,zhou2017}. Solid theoretical evidence for the existence of spin-liquid 
phases has been worked out in models with spin anisotropic super-exchange couplings, most notably the compass Kitaev model on the honeycomb lattice,
which represents a unique example of a non-trivial spin model that can be exactly solved in two spatial dimensions. Here, both gapped and gapless 
phases are present as ground states, as well as an interesting chiral state in presence of a (small) external magnetic field~\cite{kitaev2006}.
By contrast, for Heisenberg models with SU(2) spin symmetry, the situation is more debated. A predominant example, which has been intensively 
investigated in the recent past, is given by the $S=1/2$ Heisenberg model on the kagome lattice with nearest-neighbor antiferromagnetic coupling 
($J_1>0$) only. The principal motivation comes from both experimental and theoretical reasons. As far as the former ones are concerned, it is 
remarkable that different families of materials may be synthesized, providing a clean realization of this spin model (e.g., perturbations coming 
from impurities, Dzyaloshinskii-Moriya or additional inter-plane interactions are relatively small compared to the nearest-neighbor super-exchange). 
This is the case for ZnCu$_3$(OH)$_6$Cl$_2$~\cite{mendels2007,helton2007,devries2008,norman2016}, where no evidence for the insurgence of magnetic 
order is detected down to extremely small temperatures. From a theoretical point of view, the $S=1/2$ Heisenberg model on the kagome lattice 
represents one of the major challenges in quantum magnetism, given its unconventional spectrum with an exceedingly large number of low-energy 
singlet excitations~\cite{lecheminant1997,lauchli2019}.

A renewed effort to understand its physical properties followed from density-matrix renormalization group (DMRG) calculations, which suggested 
the possibility that the ground state is a so-called $\mathbb{Z}_2$ spin liquid, with topological degeneracy and a finite spin 
gap~\cite{yan2011,depenbrock2012}. An alternative scenario suggested the stabilization of a gapless spin liquid, as proposed from a parton approach 
of the original spins~\cite{hastings2000,ran2007}. Here, Abrikosov fermions are introduced to define a mean-field Hamiltonian, with $\pi$-fluxes 
piercing the hexagonal plaquettes of the lattice. As a result, Dirac points are present in the free-fermion band structure. This state has been 
dubbed U(1) {\it Dirac state}, since the leading gauge fluctuations that emerge from the mean-field Hamiltonian have U(1) symmetry. When the 
Gutzwiller projector is considered, in order to construct a suitable variational wave function for the spin model, a remarkably accurate energy 
is obtained for the nearest-neighbor Heisenberg model~\cite{iqbal2013}. In fact, further DMRG calculations with special boundary 
conditions~\cite{he2017,zhu2018} supported the existence of Dirac cones in the spinon spectrum. Furthermore, tensor networks on the infinite 
lattice~\cite{liao2017} also suggested a gapless spin liquid. This possibility immediately triggers the question of the stability of the gapless 
ground state against small perturbations. 

Here, we consider the case where a next-nearest-neighbor super-exchange coupling ($J_2$) is included, with both ferromagnetic and antiferromagnetic 
character. In recent past, only few works have investigated the nature of the ground state of the $J_1$\textendash$J_2$ Heisenberg model on the kagome 
lattice~\cite{suttner2014,iqbal2015,gong2015,kolley2015,liao2017}. For $J_2/J_1>0$ an antiferromagnetic phase with ${\bf q}=(0,0)$ pitch vector is 
expected to exist for sufficiently large values of the next-nearest-neighbor interactions; instead, for $J_2/J_1<0$ another magnetically ordered 
phase with a $\sqrt{3} \times \sqrt{3}$ pattern is present. In addition, valence-bond crystals (VBCs), with possibly large unit cells (e.g., 
containing $12$ or even $36$ sites) may represent competitive states, as suggested in previous 
works~\cite{singh2007,evenbly2010,iqbal2011,iqbal2012,huh2011,changlani2019,wietek2020}.

Within the Abrikosov-fermion approach, different variational wave functions can be defined, by allowing different terms in the fermionic state, which 
can induce the opening of a spin gap (e.g., in a $\mathbb{Z}_2$ spin liquid), the breaking of translational symmetry (leading to a VBC), or the onset 
of magnetic order. The main outcome of the present paper is that the gapless spin liquid is stable in a finite region of the $J_1$\textendash$J_2$ 
model. Indeed, for $J_2/J_1>0$, its accuracy to reproduce the exact ground-state improves with respect to the $J_2=0$ case, as indicated by a direct 
comparison with exact diagonalization on small clusters (the overlap between the gapless spin liquid and the exact ground state on $36$ sites 
increases from $0.687$ at $J_2=0$ to $0.875$ at $J_2/J_1=0.05$). Then, by increasing the ratio $J_2/J_1$, the variational wave function develops 
magnetic order with ${\bf q}=(0,0)$, namely a finite Zeeman field can be stabilized (in the thermodynamic limit) within the fermionic Hamiltonian 
(on top of the hopping pattern of the $\pi$-flux state). The transition is located at $J_2/J_1=0.11(1)$ and is weakly first order, being 
characterized by a jump in the Zeeman field. 

\begin{figure}
\includegraphics[width=\columnwidth]{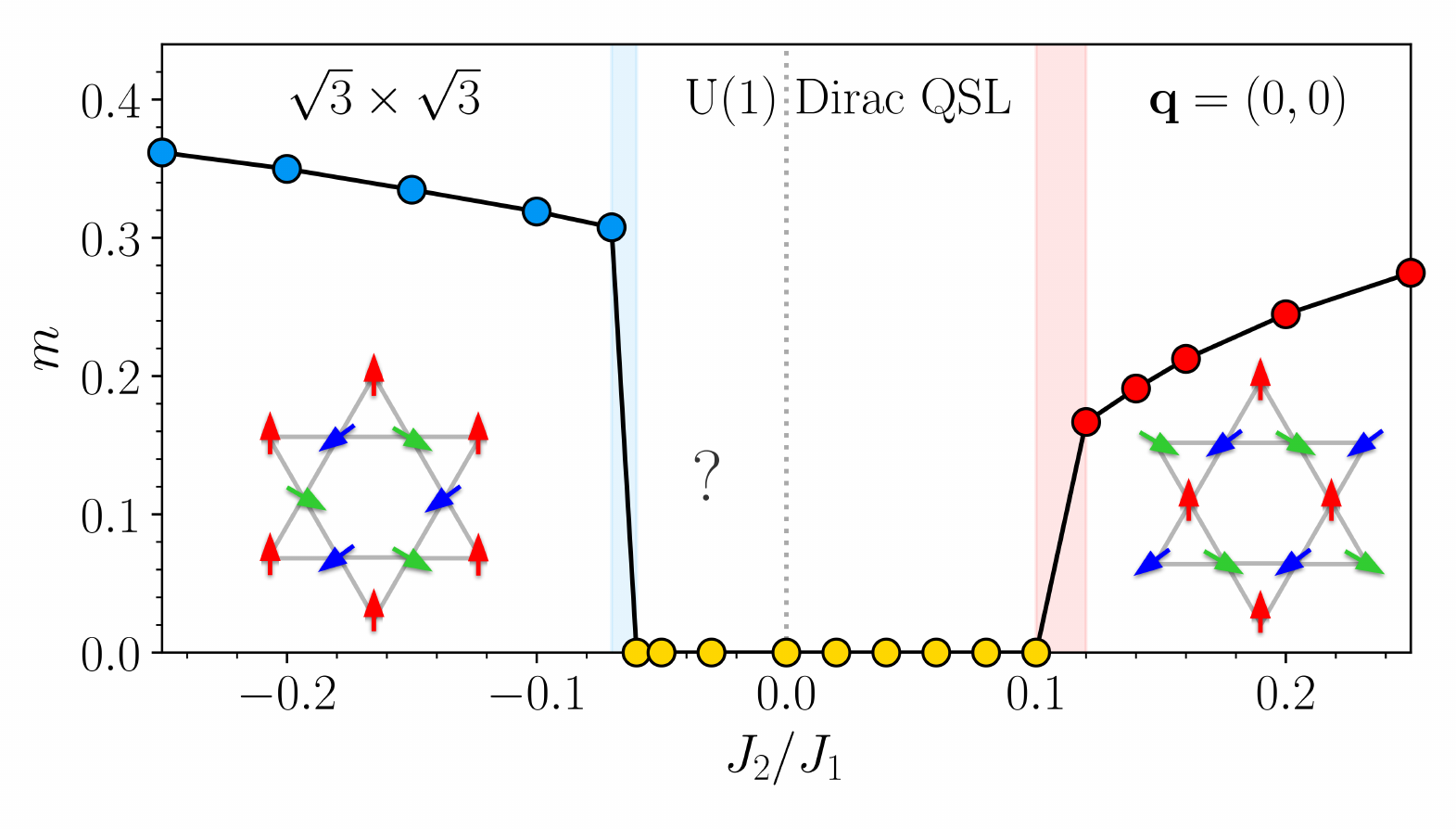}
\caption{\label{fig:QPD}
The antiferromagnetic order parameter for the $J_1$\textendash$J_2$ Heisenberg model on the kagome lattice. Different colors indicate the various 
ground-state phases: magnetic phases with ${\bf q}=(0,0)$ (red) or $\sqrt{3}\times\sqrt{3}$ periodicity (blue), and gapless spin liquid (yellow). 
The values are estimated in the thermodynamic limit (the error bars are smaller than the symbols), as shown below. The spin patterns for the two
magnetically ordered phases are also shown. The U(1) Dirac state is expected to represent the paramagnetic region for $J_2>0$; instead, for $J_2<0$
the situation is more controversial.}
\end{figure}

For $J_2/J_1<0$, the situation is more delicate. For small values of $|J_2|/J_1$ the gapless spin-liquid wave function remains stable when allowing 
additional terms in the fermionic state. Upon increasing $|J_2|/J_1$, its variational energy is overcome by a different Gutzwiller-projected state, 
with $\sqrt{3} \times \sqrt{3}$ magnetic order and hopping terms with a different flux pattern ($\pi$ flux on hexagons and up triangles, and zero 
flux on down triangles). The transition to a magnetically ordered phase is found at $J_2/J_1=-0.065(5)$. In addition, VBC states with large unit 
cells (e.g., $36$ sites) may also be stabilized and have competing energies close to the magnetic transition. The variational phase diagram, as 
obtained within our approach, is shown in Fig.~\ref{fig:QPD}. However, some care must be put on small negative values of $J_2$, where the U(1) Dirac 
spin liquid no longer represents an accurate wave function, as shown upon a comparison to the exact ground state on small clusters. This is due to 
the presence of level crossings (on $12$ sites) or avoided crossings (on $36$ sites) that happen in the low-energy singlet sector when varying 
$J_2/J_1$ close to $J_2=0$. Whether these crossings correspond to some phase transition in the thermodynamic limit is hard to resolve. Still, the 
situation remains more controversial on the $J_2<0$ side of the phase diagram, suggesting that an alternative approach may be needed when 
ferromagnetic super-exchange couplings are present.

The paper is organized as follows: in Section~\ref{sec:method}, we describe the variational method that has been used; in Section~\ref{sec:results},
we discuss our numerical results; finally, in Section~\ref{sec:concl}, we draw our conclusions.

%%%%%%%%%%%%%%%%%%%%%%%%%%%%%%%%%%%%%%%%%%%%%%%%%%%%%%%%%%%%%%%%%%%%%%%%%%%%%%%%%%%%%%%%%%%%%
%%%%%%%%%%%%%%%%%%%%%%%%%%%%%%%%%%%%%%%%%%%%%%%%%%%%%%%%%%%%%%%%%%%%%%%%%%%%%%%%%%%%%%%%%%%%%

\section{Variational wave functions}\label{sec:method}

The Abrikosov-fermion representation allows us to express the $S=1/2$ spin operators as products of fermionic creation and annihilation 
operators~\cite{baskaran1988,affleck1988,affleck1988b}:
\begin{equation}\label{eq:Sabrikosov}
\mathbf{\hat{S}}_i = \frac{1}{2} \sum_{\alpha,\beta} \hat{c}_{i,\alpha}^\dagger \boldsymbol{\sigma}_{\alpha,\beta} \hat{c}_{i,\beta}^\dagga,
\end{equation}
where $\boldsymbol{\sigma}=(\sigma_x,\sigma_y,\sigma_z)$ is a vector of Pauli matrices and $\hat{c}_{i,\alpha}^\dagger$ ($\hat{c}_{i,\alpha}^\dagga$)
creates (destroys) a fermion at site $i$ with spin $\alpha=\uparrow$, $\downarrow$. This representation fulfills the spin commutation relations, 
but enlarges the Hilbert space of the system by including unphysical states with empty or doubly-occupied sites. The variational states employed 
in this work are defined by projecting Abrikosov-fermion wave functions into the physical spin space by using the Gutzwiller projector
\begin{equation}\label{eqn:guzzo}
\hat{\mathcal{P}}_{G}=\prod_{i}(\hat{n}_{i,\uparrow}-\hat{n}_{i,\downarrow})^2,
\end{equation}
where $\hat{n}_{i,\alpha}=\hat{c}_{i,\alpha}^\dagger \hat{c}_{i,\alpha}^\dagga $. The fermionic state $|\Phi_0\rangle$ is obtained as the ground state 
of a noninteracting Hamiltonian, featuring hopping, pairing, and a fictitious Zeeman field. In particular, different symmetries of the pairing term 
can be considered, following the classifications of Ref.~\cite{lu2011}. However, the variational minimization suggests that they are not stabilized 
for the values of $J_2/J_1$ considered in this work. Then, the best wave function is obtained from a noninteracting Hamiltonian that contains only
hoppings and Zeeman fields:
\begin{equation}\label{eqn:mf-mag}
\hat{\mathcal{H}}_{\rm 0}=\sum_{(i,j),\alpha}\chi_{ij}\hat{c}_{i,\alpha}^{\dagger}\hat{c}_{j,\alpha} + h \sum_{i}\mathbf{M}_i\cdot\mathbf{\hat{S}}_i.
\end{equation}
In the following, we will include nearest- and next-nearest-neighbor hopping, $|\chi_1|=1$ (to fix the energy scale) and $\chi_2$ (as a variational
parameter), respectively; an additional parameter is the amplitude of the magnetic field $h$, while the spatial periodicity is fixed by the unit vector
$\mathbf{M}_i$, which lies in the $XY$ plane, i.e., ${\mathbf{M}_i=[\cos(\mathbf{q}\cdot{\bf R}_i+\phi_i),\sin(\mathbf{q}\cdot{\bf R}_i+\phi_i),0]}$
(where $\mathbf{q}$ is the pitch vector, ${\bf R}_i$ is the coordinate of the unit cell of site $i$, and $\phi_i$ is a sublattice-dependent angle).
The same Hamiltonian may also give rise to VBC states, by setting $h=0$ and allowing hoppings to break the space group symmetries, e.g., considering 
different values of $|\chi_1|$ and $|\chi_{2}|$ for different bonds within an enlarged unit cell~\cite{iqbal2011,iqbal2012}. Additionally, a spin-spin 
Jastrow factor is included:
\begin{equation}\label{eqn:jastrow}
\hat{\mathcal{J}}_{z}=\exp \left( \frac{1}{2}\sum_{ij} u_{ij}\hat{S}_{i}^{z}\hat{S}_{j}^{z} \right ),
\end{equation}
where $u_{ij}$ defines a set of additional variational parameters, one for each distance $|{\bf R}_i-{\bf R}_j|$. Finally, the projection 
$\hat{\mathcal{P}}_z$ onto the subspace with $\sum_i \hat{S}^z_i=0$ is also performed. As a result, the variational wave function is written as
\begin{equation}\label{eqn:physical-wf}
|\Psi_{\rm var}\rangle = \hat{\mathcal{J}}_{z} \hat{\mathcal{P}}_z \hat{\mathcal{P}}_{G}|\Phi_0\rangle.
\end{equation}
It is worth mentioning that the existence of magnetic long-range order is directly related to the presence of a finite parameter $h$ in the thermodynamic 
limit. Within magnetically ordered phases, the Jastrow factor of Eq.~\eqref{eqn:jastrow} plays an indispensable role by introducing transverse quantum 
spin fluctuations around the classical spin state~\cite{manousakis1991}. In contrast to the previous study~\cite{iqbal2015}, performed with two different
{\it Ans\"atze} for magnetic and nonmagnetic states, the present choice, based upon the noninteracting Hamiltonian~\eqref{eqn:mf-mag} allows us to have a
unique framework for these phases, also including VBC.

As previously mentioned, the U(1) Dirac spin liquid represents a very accurate variational wave function for the nearest-neighbor model ($J_2=0$). This state 
is defined by a fermionic Hamiltonian $\hat{\mathcal{H}}_{\rm 0}$ with hopping terms $\chi_1$ generating a $\pi$-flux through hexagonal plaquettes and 0-flux 
through triangles~\cite{lu2011} (an additional $\chi_2$ gives a small energy gain). For sufficiently large values of the next-neighbor super-exchange, the 
ground state acquires magnetic order, with two different pitch vectors depending on the sign of $J_2$, see Fig.~\ref{fig:QPD}. On the one hand, in the 
${\bf q}=(0,0)$ ordered phase, the optimal noninteracting Hamiltonian $\hat{\mathcal{H}}_{\rm 0}$ contains a translationally invariant magnetic field (with 
sublattice angles $\phi_i$ such as to have a relative $120\degree$ orientation between neighboring spins in the unit cell) and the same hopping structure of 
the U(1) Dirac state. On the other hand, within the $\sqrt{3} \times \sqrt{3}$ ordered phase, a magnetic unit cell of $9$ sites is required, with neighboring 
spins still having a relative $120\degree$ orientation (see Fig.~\ref{fig:QPD}). The optimal variational Ansatz is constructed from the Hamiltonian with 
a ${\bf q}=(4\pi/3a,0)$ magnetic field (where $a$ is the length of the Bravais lattice vectors) and the hopping terms of a different U(1) state, dubbed 
$[\pi,\pi]$, with $\pi$ fluxes through hexagons and up triangles (and $0$ flux through down triangles)~\cite{lu2011}. 

Our variational calculations are mostly performed on $N=3\times L \times L$ clusters, with a few exceptions (including results of Lanczos diagonalization) 
in which the tilted $N=9 \times L \times L$ clusters have been employed. All the finite-size clusters considered in this work fulfill the point group symmetries 
of the kagome lattice, and periodic boundary conditions for the Heisenberg Hamiltonian are imposed. On the contrary, the fermionic Hamiltonian~\eqref{eqn:mf-mag} 
may have periodic or antiperiodic boundary conditions along the two vectors that define the cluster. Among these four possibilities, one of them gives rise to 
zero-energy modes in the fermionic spectrum of the U(1) Dirac state. The other three choices give the same variational energy after Gutzwiller projection, 
however, each one of them breaks rotational symmetries on finite clusters~\cite{hermele2008}. Within the variational calculations, one of these three 
possibilities has been considered. To evaluate the expectation values for a given variational state, we perform a quantum Monte Carlo sampling~\cite{beccabook}. 
For the optimization of the variational parameters we make use of the stochastic reconfiguration technique~\cite{sorella2005}. 

%%%%%%%%%%%%%%%%%%%%%%%%%%%%%%%%%%%%%%%%%%%%%%%%%%%%%%%%%%%%%%%%%%%%%%%%%%%%%%%%%%%%%%%%%%%%%
\begin{figure*}
\includegraphics[width=\columnwidth]{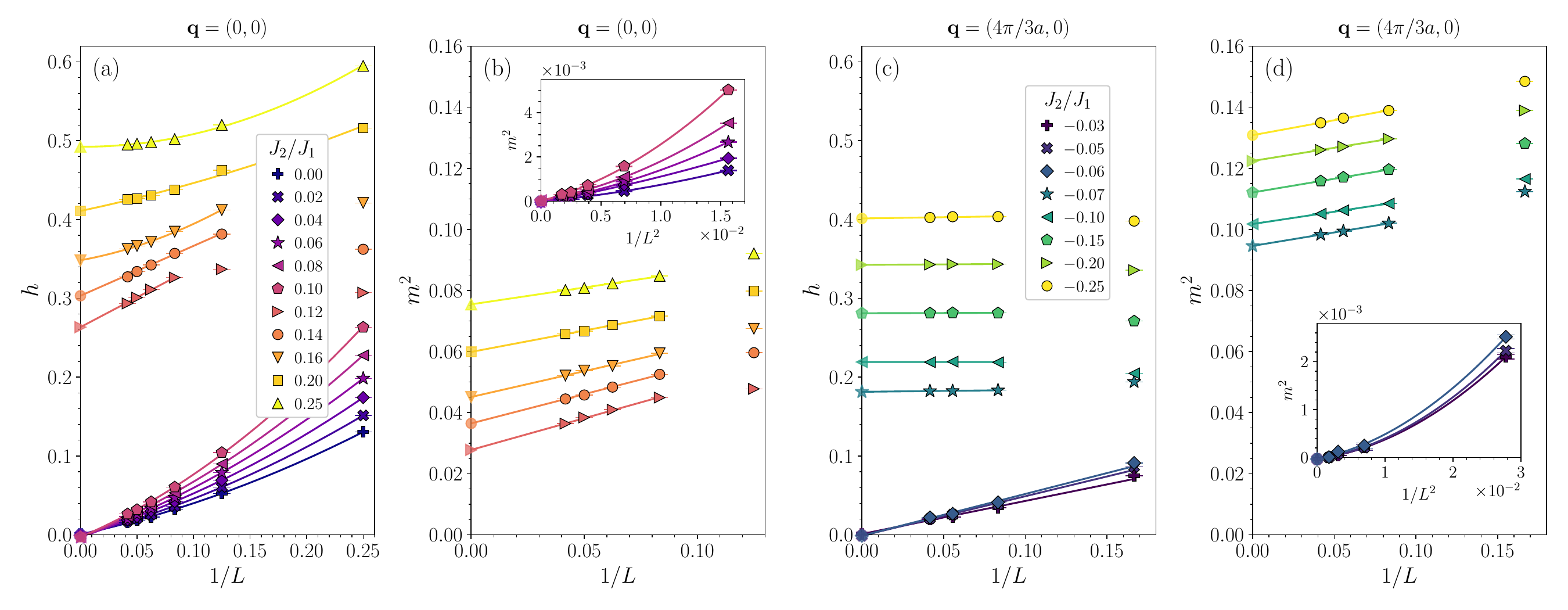}
\caption{\label{fig:fss}
Finite-size scalings of the fictitious Zeeman field $h$, see Eq.~\eqref{eqn:mf-mag}, and the square of the sublattice magnetization $m^2$ [panels 
(a) and (b) refer to the  ${\bf q}=(0,0)$ order, panels (c) and (d) refer to the $\sqrt{3}\times \sqrt{3}$ order with ${\bf q}=(4\pi/3a,0)$]. For 
$J_{2}>0$, we employed $3 \times L \times L$ clusters with $L=4n$ ($n=1,\dots,6$); for $J_{2}<0$ we used $L=6n$ ($n=1,\dots,4$). The hopping structure
of the fermionic Hamiltonian~\eqref{eqn:mf-mag} reproduces the U(1) Dirac state for $J_2/J_1 \geqslant -0.06$, while it gives the $[\pi,\pi]$ state 
for $J_2/J_1 \leqslant -0.07$. The insets in (b) and (d) show the finite-size scaling of $m^2$ within the spin-liquid regime, as a function of $1/L^2$, 
which is consistent with a power-law decay of the spin-spin correlations of the $U(1)$ Dirac state~\cite{ferrari2021}.}
\end{figure*}
%%%%%%%%%%%%%%%%%%%%%%%%%%%%%%%%%%%%%%%%%%%%%%%%%%%%%%%%%%%%%%%%%%%%%%%%%%%%%%%%%%%%%%%%%%%%%

%%%%%%%%%%%%%%%%%%%%%%%%%%%%%%%%%%%%%%%%%%%%
\floatsetup[table]{capposition=bottom}
\begin{table*}
\setlength{\tabcolsep}{8pt}
\setlength\extrarowheight{2pt}
\centering
\begin{tabular}{llllllllll}
\hline\hline
 & Phase I & Phase II & Method & $J_{2}/J_{1}$ \\
 \hline
 \multirow{9}{*}{$J_2$\textendash Antiferromagnetic}& \multirow{9}{*}{Spin liquid} & 
 \multirow{9}{*}{${\bf q}=(0,0)$}   & Variational Monte Carlo (present work)     & $0.11(1)$   \\
& &                                 & Variational Monte Carlo~\cite{tay2011}     & $0.08$      \\
& &                                 & DMRG~\cite{gong2015}                       & $0.15-0.20$ \\
& &                                 & DMRG~\cite{kolley2015}                     & $0.20$      \\
& &                                 & Tensor networks~\cite{liao2017}            & $0.045(10)$ \\
& &                                 & Exact diagonalization~\cite{changlani2018} & $0.16$      \\
& &                                 & Exact diagonalization~\cite{prelov2020}    & $0.10$      \\
& &                                 & Coupled-cluster method~\cite{richter}      & $0.127$     \\
& &                                 & One-loop PFFRG~\cite{suttner2014}          & $0.7$       \\
 \hline
 \multirow{7}{*}{$J_2$\textendash Ferromagnetic}& \multirow{7}{*}{Spin liquid} &  
 \multirow{7}{*}{${\bf q}=(4\pi/3a,0)$}          & Variational Monte Carlo (present work)     & $-0.065(5)$ \\
& &                                              & DMRG~\cite{kolley2015}                     & $-0.10$ \\ 
& &                                              & Tensor networks~\cite{liao2017}            & $-0.03$ \\
& &                                              & Exact diagonalization~\cite{changlani2018} & $-0.06$ \\
& &                                              & Exact diagonalization~\cite{prelov2020}    & $-0.10$ \\
& &                                              & Coupled-cluster method~\cite{richter}      & $-0.07$ \\
& &                                              & One-loop PFFRG~\cite{suttner2014}          & $-0.4$  \\
 \hline\hline
\end{tabular}
\caption{The value of the transition between the spin-liquid and ${\bf q}=(0,0)$ (for $J_2>0$) and $\sqrt{3} \times \sqrt{3}$ (for $J_2<0$) 
magnetic orders obtained from our present calculations, compared to different methods for the $J_1$\textendash$J_2$ Heisenberg model on the kagome lattice. 
Here, PFFRG stands for pseudo-fermion functional renormalization group.}
\label{tab:J2c}
\end{table*}
%%%%%%%%%%%%%%%%%%%%%%%%%%%%%%%%%%%%%%%%%%%%

\section{Results}\label{sec:results}

Let us now discuss our variational results. First of all, we investigate the case with antiferromagnetic next-nearest-neigbor super-exchange, i.e., 
$J_2/J_1 > 0$. Here, we consider a variational wave function that is generated from the uncorrelated Hamiltonian~\eqref{eqn:mf-mag}, including a fictitious 
Zeeman field with ${\bf q}=(0,0)$. In this regime, the best choice of the hoppings is such to obtain the U(1) Dirac state. In particular, both nearest- 
and next-nearest-neighbor hoppings are allowed~\cite{iqbal2011}, which together with the antiferromagnetic parameter $h$ and all the independent $u_{ij}$'s 
of the spin-spin Jastrow factor~\eqref{eqn:jastrow} constitute the variational parameters. After optimizing on cluster sizes up to $L=24$ (with 190
variational parameters, among which 188 for the Jastrow factor), we find that for $J_2/J_1 \geqslant 0.12$, the $h$ parameter extrapolates to a finite 
value in the thermodynamic limit [see Fig.~\ref{fig:fss}(a)], suggesting the existence of magnetic order. By contrast, for $J_2/J_1 \leqslant 0.10$, 
strong frustration is able to stabilize a quantum spin-liquid ground state; indeed, here the $h$ parameter goes to zero as $1/L^2$ for $L \to \infty$,
consistently with a power-law decay of the spin-spin correlations of the U(1) Dirac state~\cite{ferrari2021} [see Fig.~\ref{fig:fss}(a)]. We remark that,
on each finite cluster, the Jastrow factor always leads to an improvement of the variational energy, although in the spin-liquid regime the effect is 
less pronounced.

Previous studies~\cite{suttner2014,gong2015,kolley2015,liao2017,tay2011,changlani2018,prelov2020,richter} [see Table~\ref{tab:J2c}] investigated the onset 
of magnetic order for $J_2>0$. Apart from one-loop pseudo-fermion functional renormalization group calculations~\cite{suttner2014} (which are expected to 
be significantly altered at high-loop orders where convergence is reached), all other methods obtained values ranging between $J_2/J_1 \approx 0.05$ and 
$\approx 0.20$. Our estimate of the transition point lies in this range, slightly smaller than DMRG calculations~\cite{gong2015,kolley2015} but larger than 
the tensor-network evaluation~\cite{liao2017}. We mention that in a previous work of ours~\cite{iqbal2015}, we used a simplified variational wave function 
to describe the magnetic phase [including only the fictitious magnetic field $h$ but not the fermionic hopping in Eq.~\eqref{eqn:mf-mag}], leading to a 
substantial shift of the magnetic transition to larger values of $J_2/J_1$ (or, in other words, enlarging the stability region of the spin liquid by 
reducing the variational manifold of the magnetic states).

%%%%%%%%%%%%%%%%%%%%%%%%%%%%%%%%%%%%%%%%%%%%%%%%%%%%%%%%%%%%%%%%%%%%%%%%%%%%%%%%%%%%%%%%%%%%%
\begin{figure}
\includegraphics[width=\columnwidth]{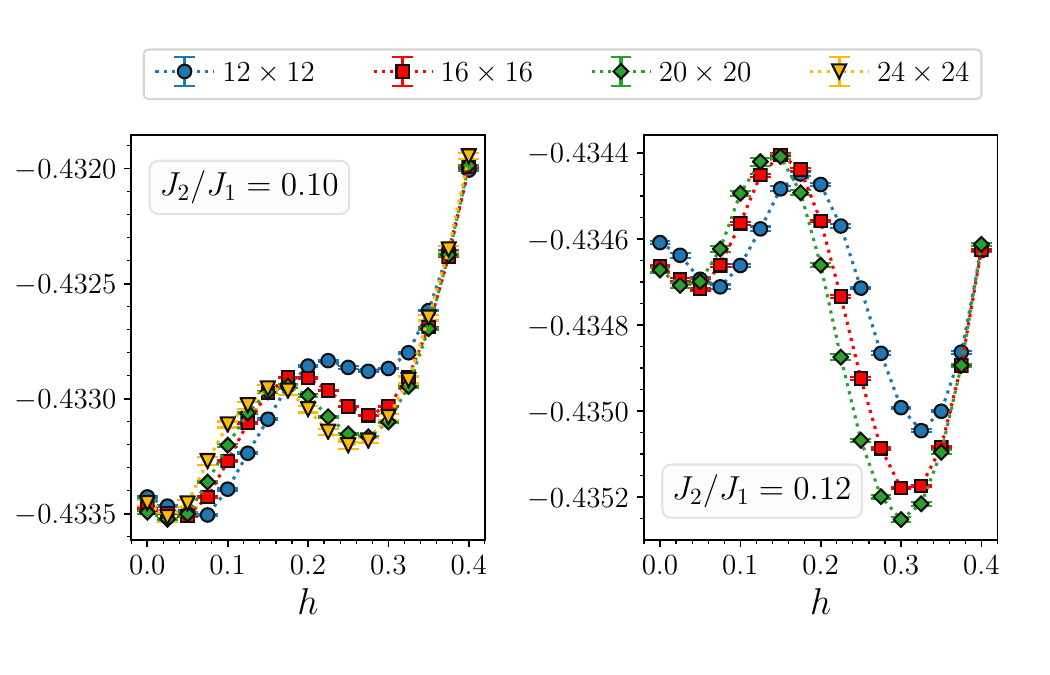}
\caption{\label{fig:landscapeAFM}
Variational energy as a function of the fictitious Zeeman field $h$. The energy landscape has been computed for $J_2/J_1=0.10$ (left panel) and 
$J_2/J_1 = 0.12$ (right panel). Clusters with $3 \times L \times L$ sites have been used, with $L = 12$, $16$, $20$, and $24$.}
\end{figure}

\begin{figure}
\includegraphics[width=\columnwidth]{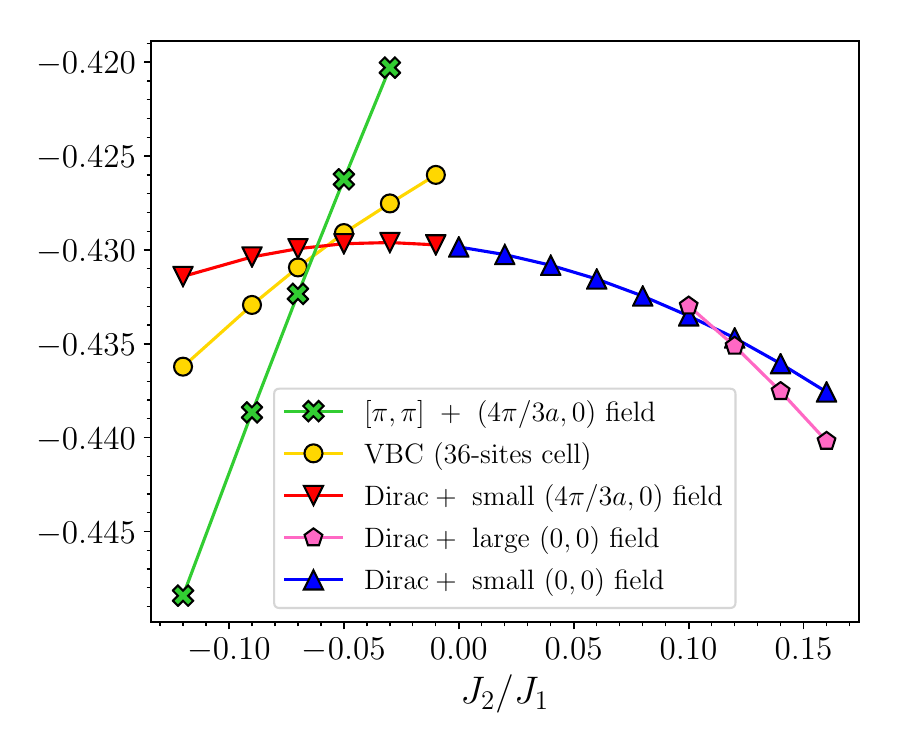}
\caption{\label{fig:energy}
Variational energies as a function of $J_2/J_1$ on the $9 \times 8 \times 8$ cluster. Different wave functions are considered, including a VBC state with 
$36$ sites in the unit cell. When two local minima in the energy landscape are present (with small and large Zeeman fileds, as in Fig.~\ref{fig:landscapeAFM}), 
both variational energies are shown. We note that the energy landscape of the $[\pi,\pi]$ state with a $(4\pi/3a,0)$ field has a single minimum as a function 
of $h$.}
\end{figure}
%%%%%%%%%%%%%%%%%%%%%%%%%%%%%%%%%%%%%%%%%%%%%%%%%%%%%%%%%%%%%%%%%%%%%%%%%%%%%%%%%%%%%%%%%%%%%

In order to have solid evidence for magnetic ordering, we compute the sublattice magnetization $m$, which is obtained from the expectation value of the 
spin-spin correlation at maximum distance (for two spins within the same sublattice):
\begin{equation}
m^2=\lim_{|i-j|\to\infty}\langle \mathbf{\hat S}_{i}\cdot \mathbf{\hat S}_{j}\rangle
\end{equation}
for the variational state $|\Psi_{\rm var}\rangle$. The magnetization displays a similar scaling as the $h$ parameter, thus confirming the extent of the 
spin-liquid regime, see Fig.~\ref{fig:fss}(b). The magnetization estimate in the thermodynamic limit is shown in Fig.~\ref{fig:QPD}. Here, we observe a 
relatively sharp jump in $m$ upon traversing the phase boundary, suggesting that the transition is not continuous. Since, a continuous transition between 
the U(1) Dirac spin liquid and the ${\bf q}=(0,0)$ state is, in principle, allowed~\cite{song2019}, we attempt to ascertain the order of the phase 
transition in our numerical simulations. For this purpose, we chart out the variational energy landscape as a function of the fictitious Zeeman field $h$: 
this is done by fixing the field $h$ to a grid of different values and optimizing only the remaining variational parameters to get the lowest energy. 
In Fig.~\ref{fig:landscapeAFM}, we show this energy landscape for different system sizes and for two values of $J_2/J_1$, one on either side of the 
transition. At $J_2/J_1=0.10$ (i.e., inside the spin-liquid regime), there are two minima: the lowest-energy one, extrapolating to $h=0$ in the 
thermodynamic limit, and another one at higher energy for finite field $h \approx 0.3$. A finite-size scaling of the energy difference between these 
minima shows that it remains finite in the thermodynamic limit. At $J_2/J_1=0.12$ (i.e., inside the magnetic regime), the two minima switch, the one 
at $h \approx 0.3$ corresponding now to the lowest energy. The energies for the two possible states (with small and large Zeeman fields) are shown
in Fig.~\ref{fig:energy} as a function of $J_2/J_1$. Hence, our variational approach clearly indicates that, in the thermodynamic limit, the best-energy 
solution has a jump from $h=0$ to a finite value for $J_2/J_1=0.11(1)$, indicative of a first-order transition.

%%%%%%%%%%%%%%%%%%%%%%%%%%%%%%%%%%%%%%%%%%%%%%%%%%%%%%%%%%%%%%%%%%%%%%%%%%%%%%%%%%%%%%%%%%%%%
\begin{figure}
\includegraphics[width=\columnwidth]{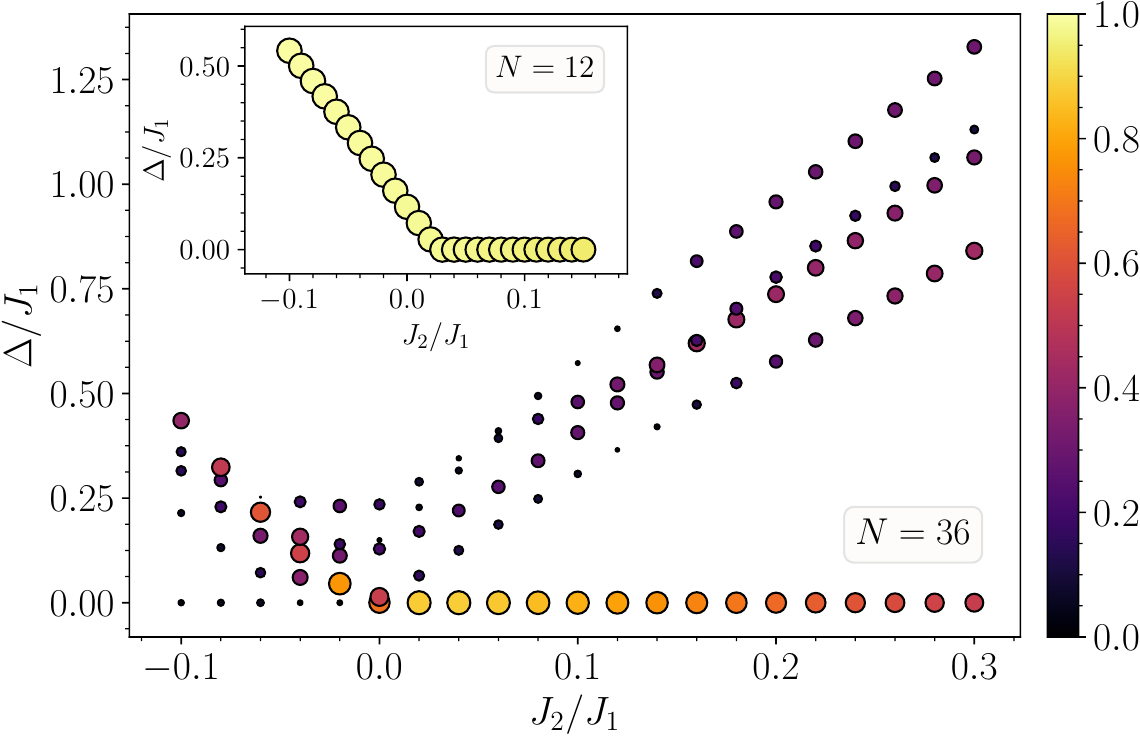}
\caption{\label{fig:overlap}
Overlap between the symmetrized U(1) Dirac state $|\Psi_{\rm sym}\rangle$ and few low-energy exact eigenstates on the $36$-site cluster, obtained by Lanczos 
diagonalization. Both the area and the colors of the circles represent the value of the overlap. On the horizontal axis we report the value of $J_2/J_1$, 
while on the vertical axis we show the energy gap $\Delta$ of the exact eigenstates with respect to the ground state, in units of $J_1$. Notice that, the 
variational Ansatz $|\Psi_{\rm sym}\rangle$ has a finite overlap only with the exact eigenstates belonging to the same symmetry sector. In the inset, analogous 
results on the $12$-site cluster are reported.}
\end{figure}

\begin{figure}
\includegraphics[width=\columnwidth]{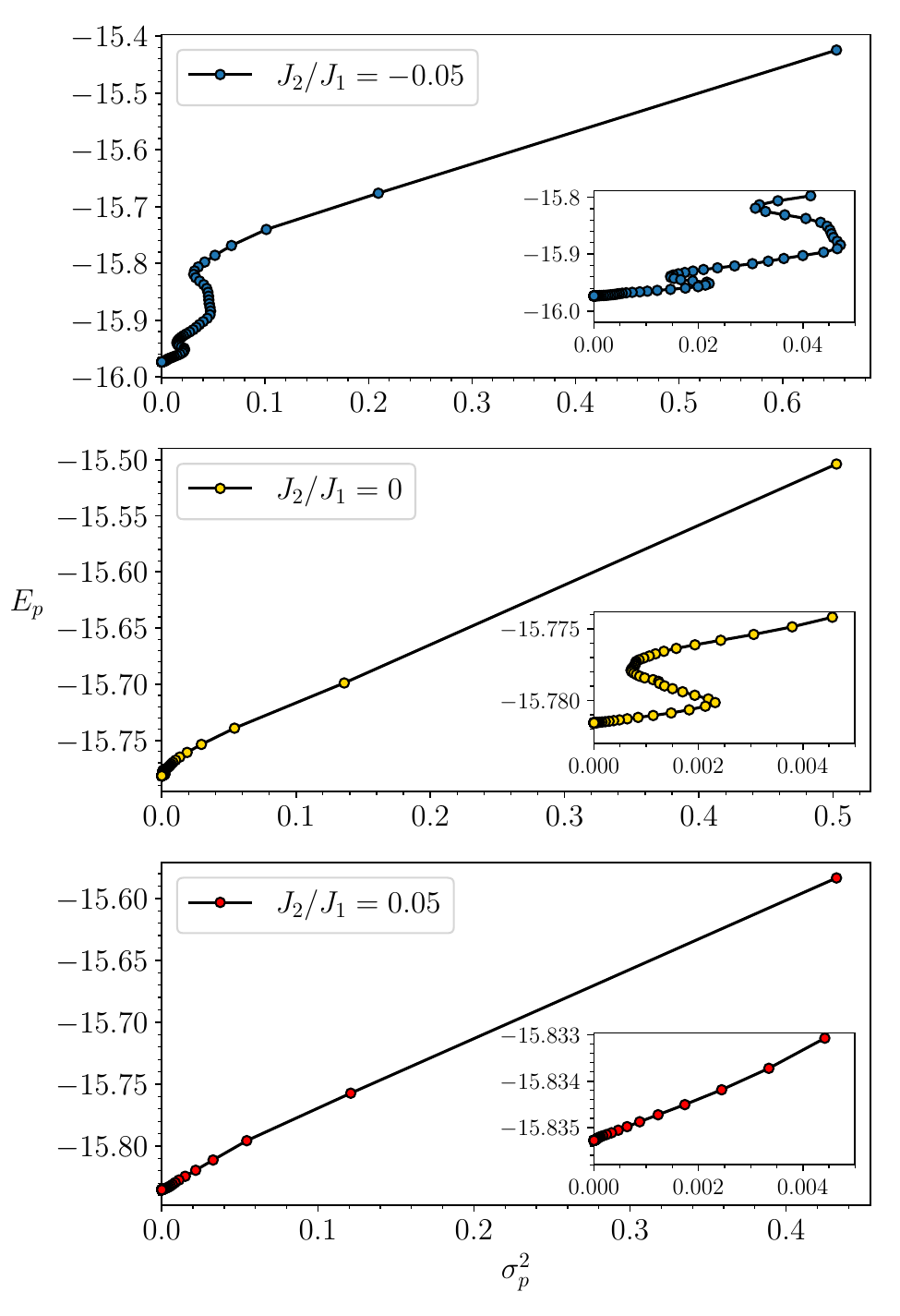}
\caption{\label{fig:lanczos}
Energy $E_p$ [see Eq.~\eqref{eqn:energy}] versus variance $\sigma^2_p$ [see Eq.~\eqref{eqn:variance}] starting from the initial state given by the 
symmetrized U(1) Dirac Ansatz for three values of the ratio $J_2/J_1$. The energy (variance) is given in units of $J_1$ ($J_1^2$). Calculations are done 
for a cluster with $36$ sites.}
\end{figure}
%%%%%%%%%%%%%%%%%%%%%%%%%%%%%%%%%%%%%%%%%%%%%%%%%%%%%%%%%%%%%%%%%%%%%%%%%%%%%%%%%%%%%%%%%%%%%

Let us now discuss the accuracy of the variational wave function when compared to the exact ground state for small finite-size lattices, e.g., the $36$ 
sites cluster (namely, $9 \times 2 \times 2$, still possessing all the symmetries of the infinite kagome lattice). Here, in order to make a neat 
comparison with the exact ground state, we construct a fully-symmetric U(1) Dirac state (taking nearest-neighbor hoppings only). In fact, even though in 
Ref.~\cite{hermele2008} it was claimed that this is not possible on $36$ sites, we verified that a suitable linear combination of the three possible 
choices of boundary conditions that do not give zero-energy modes corresponds to a state, $|\Psi_{\rm sym}\rangle$, which lies in the same symmetry 
subspace of the exact ground state, $|\Psi_{\rm ex}\rangle$. First of all, we compute the overlap between $|\Psi_{\rm sym}\rangle$ with the first few 
exact eigenstates in the same symmetry subspace, as a function of $J_2/J_1$. The results are shown in Fig.~\ref{fig:overlap}, where the case of a small 
cluster with $12$ sites is also reported. Remarkably, the overlap with the exact ground state increases when going from $J_2=0$ up to $J_2/J_1 \approx 0.05$.

Further highly convincing evidence that the ground-state wave function is well approximated by the U(1) Dirac spin-liquid Ansatz comes from performing 
the Lanczos technique, which allows us to obtain the exact ground state on a small cluster by an iterative procedure~\cite{laflorencie2004}. Starting 
from an arbitrary quantum state $|\Psi_0\rangle$, after $p$ iterations, an estimate of the ground state is given by
\begin{equation}
|\Psi_p\rangle = \left ( \sum_{k=0}^p \alpha_k \hat{\mathcal{H}}^k \right )|\Psi_0\rangle,
\end{equation}
where the coefficients $\alpha_k$ are found by minimizing the energy
\begin{equation}\label{eqn:energy}
E_p = \langle \Psi_p|\hat{\mathcal{H}}|\Psi_p\rangle.
\end{equation}
Here, we choose $|\Psi_0\rangle \equiv |\Psi_{\rm sym}\rangle$ and compute the energy $E_p$ as a function of the variance
\begin{equation}\label{eqn:variance}
\sigma^2_p = \langle \Psi_p|\hat{\mathcal{H}}^2|\Psi_p\rangle - \langle \Psi_p|\hat{\mathcal{H}}|\Psi_p\rangle^2,
\end{equation}
which tends to zero when $p \to \infty$. The results are shown in Fig.~\ref{fig:lanczos} for different values of $J_2/J_1$. For both $J_2=0$ and $0.05$,
an approximately linear behavior $E_p \approx E_{\rm ex} + {\rm const} \times \sigma^2_p$ is observed, suggesting a smooth convergence of the initial
wave function to the exact ground state. Indeed, an extrapolation of the total energy based on the first three steps of the Lanczos procedure ($p=0,1,2$) 
gives $E/J_1 \approx -15.769$, compared to the exact value $E_{\rm ex}/J_1=-15.781$, for $J_2=0$. Similar results are also obtained for $J_2/J_1=0.05$, 
i.e., $E/J_1 \approx -15.826$ compared to $E_{\rm ex}/J_1=-15.835$. These results confirm the ones reported in Fig.~\ref{fig:overlap}, showing that the
variational wave function has a large overlap with the exact ground state (for these values of $J_2/J_1$). Therefore, we are confident that the U(1) 
Dirac state faithfully represents the exact ground state of the Heisenberg model on the kagome lattice, especially in presence of a small antiferromagnetic 
$J_2/J_1$.

Then, we move towards investigating the regime with ferromagnetic $J_2$, i.e., $J_2/J_1<0$. Here, we fix ${\bf q}=(4\pi/3a,0)$ in the fermionic 
Hamiltonian~\eqref{eqn:mf-mag}. In Fig.~\ref{fig:energy}, we compare the energies for different wave functions, corresponding to local minima in the 
variational energy landscape. While for $J_2/J_1 \geqslant -0.06$, the best Ansatz is given by the U(1) Dirac state with a small $h$ parameter 
[eventually extrapolating to zero in the thermodynamic limit, see Fig.~\ref{fig:fss}(c)], for $J_2/J_1 \leqslant -0.07$, the best state is magnetically
ordered and obtained by employing a different flux pattern, i.e., the $[\pi,\pi]$ state defined in Ref.~\cite{lu2011}. Therefore, a first-order transition 
is expected. A detailed size scaling of the magnetization is reported in Fig.~\ref{fig:fss}(d), confirming the existence of a magnetic state for 
$J_2/J_1 \leqslant -0.07$. Our estimate of the phase boundary is in good agreement with those from other methods as shown in Table~\ref{tab:J2c}. 
In previous works that used similar {\it Ans\"atze} for the ground-state wave function~\cite{iqbal2011,iqbal2012}, we proposed that the U(1) Dirac spin 
liquid should give way to a $36$-site VBC for $J_2/J_1 \approx -0.045$; however, the present results, with magnetic ordering emerging for 
$J_2/J_1 \lesssim -0.06$, suggest that a phase with VBC order is highly unlikely, or may be stabilized only in a sliver of parameter space close to the 
$\sqrt{3} \times \sqrt{3}$ magnetic ordered region, see Fig.~\ref{fig:energy}. Indeed, in this regime, the variational energies of the VBC candidates 
are similar to the ones of other competing states, i.e., the U(1) Dirac state and the magnetic one~\cite{changlani2019}.

In spite of these variational results, we must emphasize that the accuracy of the spin-liquid wave function strongly deteriorates as soon as a small 
ferromagnetic $J_2$ is included. Indeed, the overlap of the symmetrized U(1) Dirac state with the exact ground state is very small, as shown in 
Fig.~\ref{fig:overlap} for a $36$-site cluster. Rather, the U(1) Dirac state has a significant overlap with an exact excited eigenstate. The Lanczos 
procedure also confirms that this variational state is not smoothly connected to the exact ground state, since the linear extrapolation converges to
an energy that is well above the ground-state one, see Fig.~\ref{fig:lanczos}. In fact, on the $36$-site cluster there is an avoided crossing at 
$J_2/J_1 \approx 0$ (in the fully-symmetric subspace)~\cite{li2021}: the lowest-energy state for $J_2>0$ is well described by the U(1) Dirac spin 
liquid, while the one for $J_2<0$ is not. On $12$ sites, a similar behavior is observed, with an actual level crossing for $J_2$ slightly larger than 
$0$. These results put some doubts into the variational outcomes, for which the Dirac state remains stable up to $J_2/J_1 \approx -0.06$, with no other 
wave functions that we are able to construct, within the present Gutzwiller-projected states, having a lower energy. Indeed, we verified that both 
symmetric and lattice nematic $\mathbb{Z}_{2}$~\cite{lu2011} as well as chiral U(1) and $\mathbb{Z}_2$~\cite{bieri2016} states cannot be stabilized 
(or are not energetically competing with the Dirac state). In addition, a few VBCs with $12$-site unit cell (of the diamond type~\cite{yan2011,wietek2020}) 
have been assessed, with no gain in energy.

\section{Conclusions}\label{sec:concl}

In this work, we analyzed the $S=1/2$ $J_1$\textendash$J_2$ Heisenberg model by using a family of variational wave functions constructed from Abrikosov 
fermions, able to describe both spin liquid and magnetic phases {\it on the same footing}. This approach was previously shown to be successful for the 
case with nearest-neighbor interactions only~\cite{iqbal2013}. Here, we provided evidence that, for antiferromagnetic values of the next-nearest-neighbor 
super-exchange, the U(1) Dirac state remains stable up to $J_2/J_1=0.11(1)$; then, for larger values of $J_2/J_1$ a magnetically ordered state settles 
down, with ${\bf q}=(0,0)$ pitch vector, in agreement with other numerical methods~\cite{gong2015,kolley2015,liao2017,tay2011,changlani2018,prelov2020,richter}. 
Note that, although a first order transition is found, a continuous transition between the Dirac state and the ${\bf q}=(0,0)$ magnetic phase is not 
forbidden~\cite{song2019}. Within the (gapless) spin-liquid regime, no energy gain is obtained by allowing pairing terms in the noninteracting fermionic 
Hamiltonian~\eqref{eqn:mf-mag}, in analogy to what has been emphasized for the case with $J_2=0$~\cite{iqbal2011b,iqbal2018}. In addition, no VBC order 
has been detected by allowing nonuniform hopping amplitudes. The fact that the U(1) Dirac state faithfully represents the exact ground state of the 
$J_1$\textendash$J_2$ model for $0 \lesssim J_2/J_1 \lesssim 0.10$ also follows from a direct comparison with exact calculations on small clusters. 
For example, on the $36$-site cluster, a linear combination of Dirac states with three different boundary conditions can be constructed to have all the 
symmetries of the infinite lattice. This variational state (with no adjustable variational parameters) has quite a large overlap with the exact ground state, 
e.g., $0.875$ for $J_2/J_1=0.05$.

By contrast, the ferromagnetic regime, i.e., $J_2/J_1<0$, is more problematic and asks for future investigations. Indeed, the U(1) Dirac state continues to 
give the lowest energy within Gutzwiller-projected fermionic {\it Ans\"atze} up to $J_2/J_1=-0.065(5)$, where a magnetic state, with $\sqrt{3} \times \sqrt{3}$ 
periodicity, becomes the best variational wave function. No signal for opening of a spin gap has been detected for $-0.06 \lesssim J_2/J_1 \lesssim 0$, 
including the instability towards symmetric and lattice nematic $\mathbb{Z}_2$ spin liquids~\cite{lu2011}, U(1) and $\mathbb{Z}_{2}$ chiral spin 
liquids~\cite{bieri2016}, or VBCs with different unit cells (most notably containing $12$ or $36$ sites)~\cite{iqbal2011,iqbal2012}. Nonetheless, a comparison 
with exact calculations on small sizes shows that the U(1) Dirac state no longer accurately represents the ground state of the $J_1$\textendash$J_2$ model. 
For example, on $36$ sites for $J_2/J_1=-0.05$, the overlap between the (symmetrized) spin-liquid state and the exact ground state is only $0.063$. This fact
roots itself in the existence of an avoided crossing that changes the nature of the ground-state wave function. The resulting lowest-energy state does not 
seem to be described by any simple Gutzwiller-projected fermionic Ansatz. Whether this change in the low-energy sector is relevant for the true thermodynamic 
limit, or is only peculiar to the $36$-site cluster is hard to ascertain. Certainly, the $J_2=0$ point is at the crossroads between different quantum phases 
(including the gapless spin liquid and magnetic phases, but possibly also VBC and chiral states, or even more exotic possibilities), and a crucial question 
to address in the future is the character of the several singlet states which populate the low-energy spectrum~\cite{yao2020}.

\section{Acknowledgments}

We thank A. Paramekanti, A. Vishwanath, A. Wietek, H. Changlani, S. Pujari, S. Sachdev, S.-S. Gong, and Y.-C. He for helpful discussions. We acknowledge the kind hospitality of the Centro de Ciencias de Benasque Pedro Pascual, Benasque, Spain, during the workshop “Entanglement in Strongly Correlated Systems” (2020) where this project was initiated. Y.I. acknowledges financial support by SERB, Department of Science and Technology (DST), India through the Startup Research Grant No.~SRG/2019/000056, MATRICS Grant No.~MTR/2019/001042, and the Indo-French Centre for the Promotion of Advanced Research (CEFIPRA) Project 
No. 64T3-1. This research was supported in part by the National Science Foundation under Grant No.~NSF~PHY-1748958, the Abdus Salam International Centre for Theoretical Physics (ICTP) through the Simons Associateship scheme funded by the Simons Foundation, IIT Madras through the Institute of Eminence (IoE) program for establishing the QuCenDiEM group (Project No. SB20210813PHMHRD002720) and FORG group (Project No. SB20210822PHMHRD008268), and the International Centre for Theoretical Sciences (ICTS), Bengaluru, India during a visit for participating in the program Novel phases of quantum matter (Code: ICTS/topmatter2019/12). Y.I. acknowledges the use of the computing resources at HPCE, IIT Madras. F.F. acknowledges support from the Alexander von Humboldt Foundation through a postdoctoral Humboldt fellowship. D. P. acknowledges support by the TNSTRONG ANR-16-CE30-0025 and TNTOP ANR-18-CE30-0026-01 grants awarded by the French Research Council.

\end{document}